\documentclass{article}

\def\xxinput#1{\input#1}

\xxinput{vsolj02.sty}

\usepackage[dvipdfmx]{graphicx}
\usepackage[comma,colon]{natbib}

\usepackage[OT2,T1]{fontenc}

\def\cite{\citealt}
\setcitestyle{aysep={}}

\newcounter{author}
\def\altaffilmark#1{$^{#1}$}
\def\altaffiltext#1{$^{#1}$\,}

\def\authorcount#1#2{{\refstepcounter{author}\label{#1}
                     \altaffiltext{\ref{#1}}{#2}}}

\begin{document}

\begin{center}

\title{SU UMa-type supercycle in the IW And-type dwarf nova BO Cet}
\vskip -2mm
\title{above the period gap}

\author{
        Taichi~Kato\altaffilmark{\ref{affil:Kyoto}}
}
\email{tkato@kusastro.kyoto-u.ac.jp}

\authorcount{affil:Kyoto}{
     Department of Astronomy, Kyoto University, Sakyo-ku,
     Kyoto 606-8502, Japan}

\end{center}

\begin{abstract}
\xxinput{abst.inc}
\end{abstract}

\section{Introduction}

   Major categories of cataclysmic variables (CVs) include
novalike (NL) variables and dwarf novae (DNe).
In NL-type variables, the accretion disk is thermally
stable due to the high mass-transfer rate from the secondary.
In DNe, the disk is thermally unstable and shows outbursts.
[For general information of CVs, see e.g., \citet{war95book}].
DNe are subdivided (mainly) into SS Cyg stars, Z Cam stars
and SU UMa stars.  Z Cam stars show standstills (equivalent
to NL-type thermally stable states) and SU UMa stars show
superoutbursts in addition to ordinary outbursts in SS Cyg stars.
During superoutbursts, superhumps with periods a few percent
longer than the orbital periods are observed, and they are
considered as the defining characteristics of SU UMa stars
(see e.g., \cite{osa96review}).
These superhumps are considered to be the result of
the 3:1 resonance in the disk \citep{whi88tidal,hir90SHexcess,lub91SHa}.
Only systems with relatively small mass ratios ($q=M_2/M_1$)
are considered to hold a large disk reaching the radius
of the 3:1 resonance and this is considered to determine
whether a certain system can show an SU UMa-type phenomenon.
The borderline of $q$ for SU UMa stars is usually considered
somewhere between 0.25 \citep{whi88tidal} and
0.33 \citep{mur00SHintermediateq}.

   Recently, some of Z Cam stars are known to show atypical
behavior (see also \cite{ham14zcam}), often characterized by
(1) slowly rising standstills, sometimes with damping oscillations,
are terminated by brightening,
(2) the sequence is often very regular with almost a constant
recurrence time and (3) deep dips are sometimes seen following
brightening \citep{kat19iwandtype}.
These objects are called IW And stars [Note that not all items
are always seen in all IW And stars.  The item (1) can be
considered as the defining characteristic].  There is a suggestion
that the IW And-type phenomenon is a manifestation of
a limit cycle reflecting the increase of the angular momentum
of the disk during a sequence of small outbursts in
the outer part of the disk
\citep{kim20kic9406652,kim20iwandmodel}.
This picture is very similar to SU UMa stars having two
types of cycles (outburst cycle and supercycle)
\citep{osa89suuma,osa96review} and led to a suggestion
that tidal truncation might work in IW And stars just
as the 3:1 resonance works in SU UMa stars [subsection 4.1
in \citet{kim20kic9406652} and \citet{kat21bocet}].

   BO Cet is the first object showing both characteristics
of SU UMa stars and Z Cam/IW And stars \citep{kat21bocet}
[currently two more (but much fainter) objects are known:
MGAB-V349 (vsnet-alert 26832)\footnote{
  $<$http://ooruri.kusastro.kyoto-u.ac.jp/mailarchive/vsnet-alert/26832$>$.
} and
ZTF J181732.64$+$101954.5 (vsnet-alert 26776)\footnote{
  $<$http://ooruri.kusastro.kyoto-u.ac.jp/mailarchive/vsnet-alert/26776$>$.
}]
and this object is expected to provide a clue in understanding
the relation between these two classes of DNe.

\section{BO Cet}

   BO Cet was initially discovered as a NL-type CV
by R. Remillard (1992) but no details were published.
The CV nature was confirmed by \citet{rod07newswsex}.
The orbital period was initially detected by
the members of the Center for Backyard
Astrophysics.\footnote{
   J. Patterson in 2002,
$<$http://cbastro.org/communications/news/messages/0274.html$>$.
}
and was refined by \citet{bru17CVphot1}, placing BO Cet
above the period gap, which appeared to be consistent with
the NL-type classification.

   This object, however, was confirmed to be a dwarf nova
by \citet{kat21bocet}.  Using the data by
the VSNET Collaboration \citep{VSNET}, \citet{kat21bocet}
clarified that the object is an eclipsing object
with an orbital period of 0.139835~d.  Furthermore,
\citet{kat21bocet} showed that the object also showed
IW And-type standstills.
The most unexpected finding by \citet{kat21bocet} was
that BO Cet showed superhumps during one of its long outburst.
Superhumps had a period 7.8\% longer than the orbital period
leading to an estimation of the mass ratio of $q$=0.31--0.34.
Superhumps during a long outburst defines the SU UMa-type
dwarf nova and it was very unusual that an object far above
the period gap showed the SU UMa-type phenomenon
(but not unprecedented).
It was considered that $q$ of BO Cet is very close to
the upper limit of $q$ in which the disk can reach
the radius of the 3:1 resonance.
BO Cet could have accidentally reached this radius,
since there had been no such a phenomenon recorded in
this object despite past observations for more than 10 years,
and this might implicitly suggest such a superoutburst should
be rare.  This suggestion, however, was disproven only two
years later.

\section{Observations and Analysis}\label{sec:obs}

\begin{figure*}
\begin{center}
\includegraphics[width=16cm]{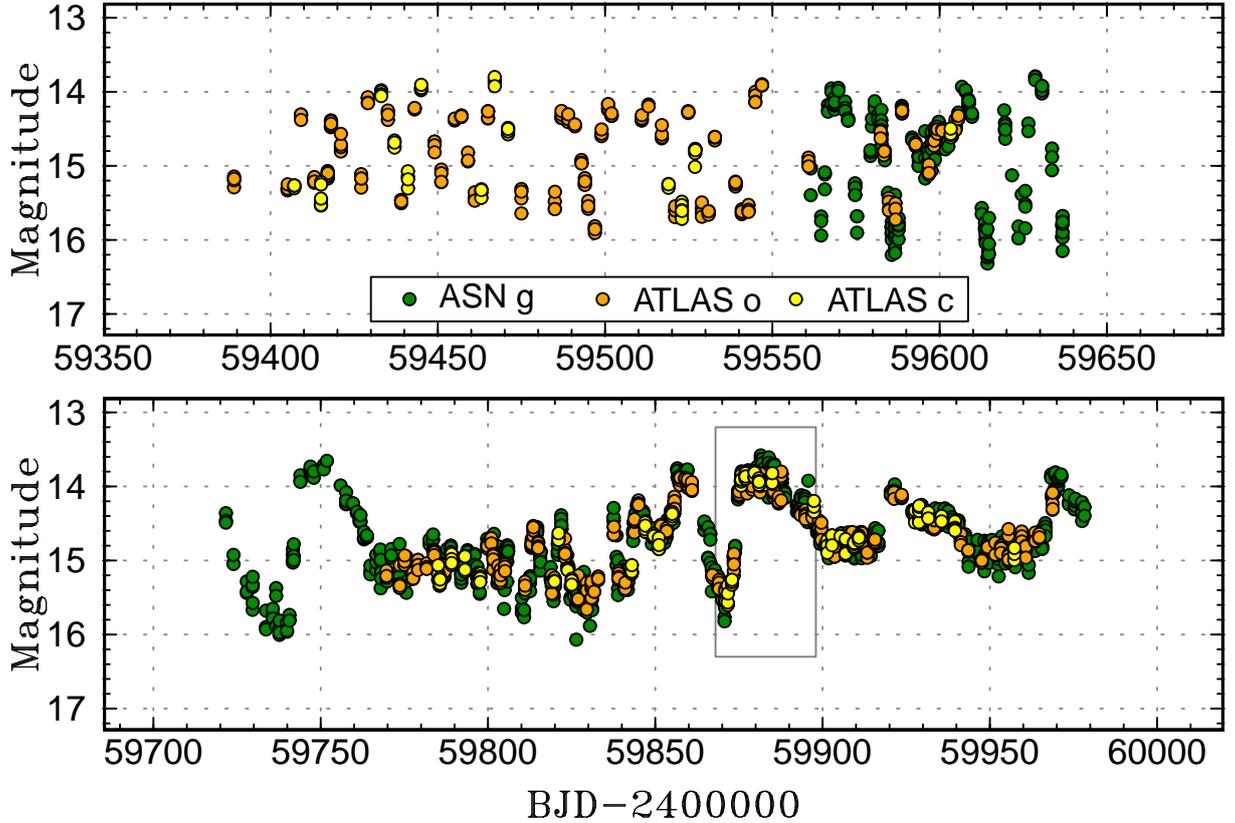}
\caption{
   Light curve of BO Cet in 2021--2023.
   ASAS-SN (ASN) and ATLAS observations are used.
   In the upper panel, an IW And-type standstill (terminal
   brightening with sudden fading) was recorded
   (BJD 2459593--2459610, 2022 January).
   In the lower panel, an SU UMa-like supercycle was recorded
   between BJD 2459744 and 2459890: following a long, bright
   outburst, a sequence of short repeated outbursts with increasing
   amplitudes led to a long, bright outburst. 
   The long, bright outburst starting on BJD 2459875
   was confirmed to be a superoutburst (enlargement of the grey box
   in figure \ref{fig:lcso}).
}
\label{fig:lc}
\end{center}
\end{figure*}

\begin{figure*}
\begin{center}
\includegraphics[width=16cm]{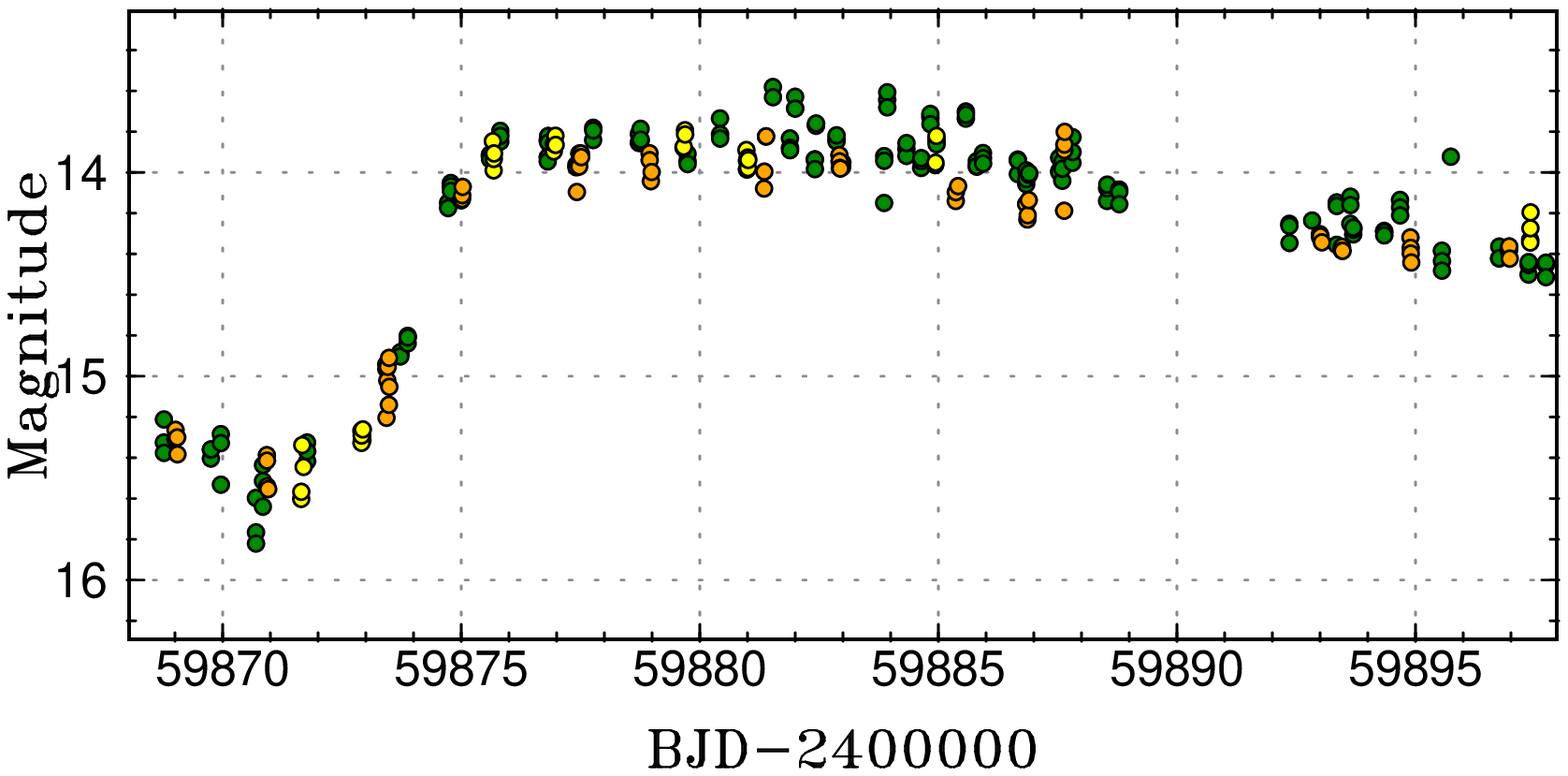}
\caption{
   Enlargement of the light curve of the 2022 superoutburst.
   The symbols are the same as in figure \ref{fig:lc}.
   The scatter in the middle part of the light curve indicates
   superhumps.
}
\label{fig:lcso}
\end{center}
\end{figure*}

   Using the All-Sky Automated Survey for Supernovae
(ASAS-SN) Sky Patrol data \citep{ASASSN,koc17ASASSNLC} and
the Asteroid Terrestrial-impact Last Alert System
(ATLAS: \cite{ATLAS}) forced photometry \citep{shi21ALTASforced},
I noticed that BO Cet gradually brightened following a bright
outburst in 2022 June.  During this brightening phase,
small outbursts with increasing amplitudes were recorded
(figure \ref{fig:lc} lower panel: BJD 2459765--2459850).
The object stopped fading and brightened (BJD 2459850--2459861)
just like an IW And-type standstill.  After reaching a minimum
around BJD 2458871, this object underwent a long, bright
outburst (figures \ref{fig:lc}, \ref{fig:lcso}).

\begin{figure*}
\begin{center}
\includegraphics[width=14cm]{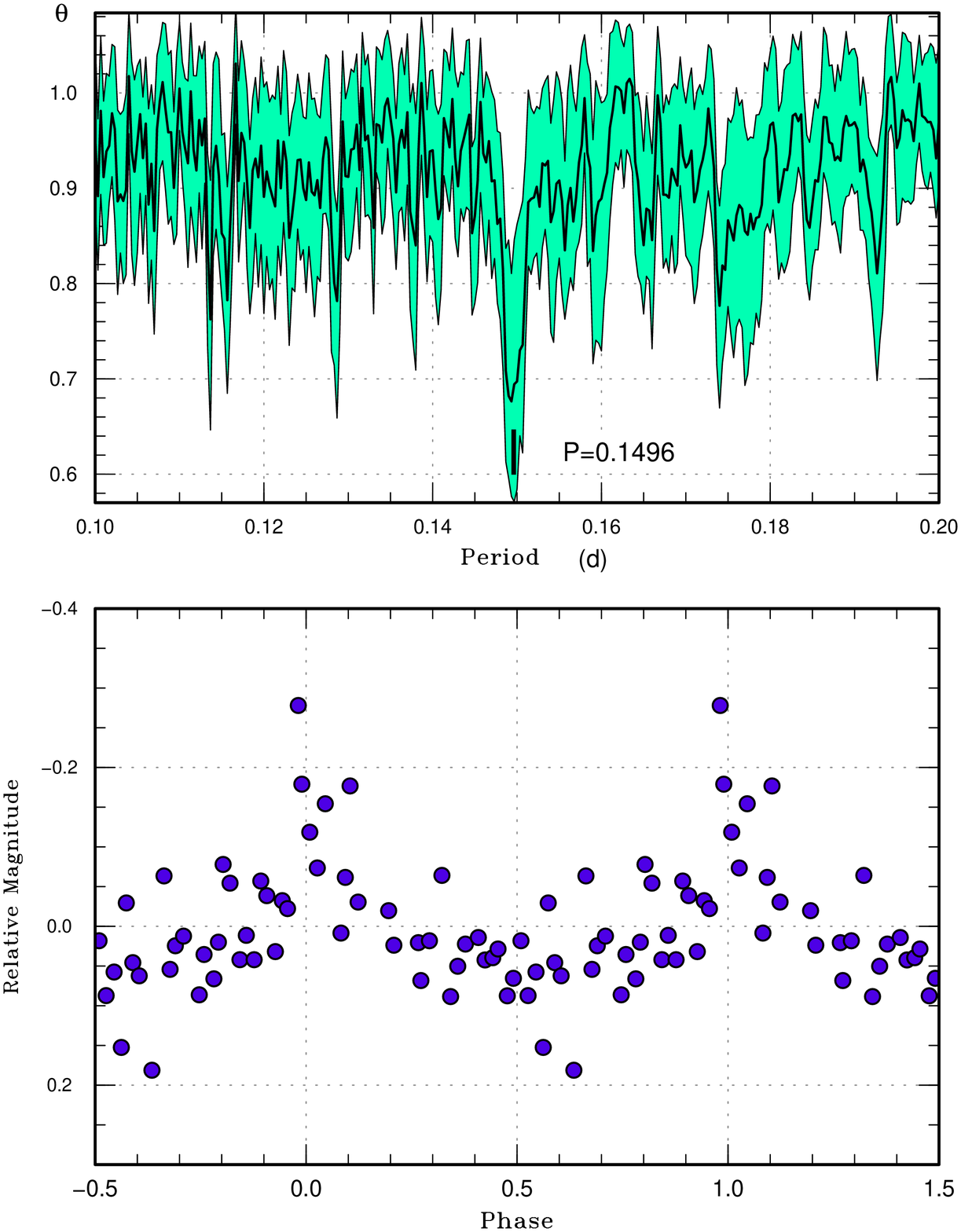}
\caption{
   Superhumps of BO Cet in 2022 (interval BJD 2459876.8--2459887.9).
   (Upper): PDM analysis.  The bootstrap result using
   randomly contain 50\% of observations is shown as
   a form of 90\% confidence intervals in the resultant 
   $\theta$ statistics.
   (Lower): Phase plot.
}
\label{fig:shpdm}
\end{center}
\end{figure*}

   During this long outburst, the increasing scatter in
the light curve was suggestive of developing superhumps
(figure \ref{fig:lcso}).  A phase dispersion minimization
(PDM; \cite{PDM}) analysis indeed detected a strong
superhump signal (figure \ref{fig:shpdm}).
I used the methods of \citet{fer89error}
and \citet{Pdot2} to determine 1$\sigma$ errors.
The resultant period of 0.1496(6)~d is in good agreement
with the value 0.15069(3)~d obtained during the 2020
superoutburst \citep{kat21bocet}.  No significant superhump
signal was detected after BJD 2459888.  This was probably
due to the decrease in the amplitude of superhumps
and the limited number of observations (10 per day or
even less).  If a time-resolved photometric campaign
had been conducted, the superhump signal may have been
detected for a longer time.

\section{Discussion}

\subsection{Supercycle}

   This observation for the first time detected a typical
supercycle for an SU UMa star in BO Cet: repeated short normal outbursts
with increasing amplitudes between two long outbursts.
The second long outburst was confirmed to be a superoutburst.
Although the nature of the first long outburst was unknown due to
the lack of observations, it could also have been a superoutburst
as seen from the shape similar to the second one.
Although the presence of a supercycle is not surprising
for an SU UMa star, it was not evident in 2020 when
the first superoutburst was detected.  This observation strongly
supports the idea that the accumulating mass and angular momentum
in the disk during repeated normal outbursts caused the disk
to expand to the radius of the 3:1 resonance
\citep{osa89suuma,osa96review}.

\subsection{Relation to IW And-type phenomenon}

   As shown in section \ref{sec:obs}, the final outburst
before the superoutburst had characteristics similar to
an IW And-type standstill.  This finding supports
the interpretation that terminal brightening of
an IW And-type standstill occurs when the disk radius
reaches a certain upper limit \citep{kim20kic9406652,kat21bocet},
possibly the radius of tidal truncation.

   The 2022 phenomenon of BO Cet was different from
an IW And-type standstill in that it showed increasing
amplitudes of dwarf-nova outbursts while damping oscillations
are seen in IW And stars.  This difference may have been
due to the difference in the initial condition when
the cycle started following a long outburst: in 2022,
the cycle started when the cool region in the disk in BO Cet
was larger than in the IW And state (at the start of the cycle,
BO Cet was 0.5~mag fainter than average standstills
of the same object).

\subsection{Failed superoutburst?}

   As seen in several SU UMa stars (usually with long
orbital periods) near the stability border of
the 3:1 resonance, an outburst just preceding a superoutburst
tends to be long and bright.  Such an outburst is
possibly a failed superoutburst, in which the duration
of the outburst is not long enough for the 3:1 resonance
to develop despite that the radius already reaches the radius
of the 3:1 resonance.
The case in BO Cet in 2022 might be the same.
This final outburst also corresponded to IW And-type
terminal brightening.  In a case in which the radii of
the 3:1 resonance and tidal truncation are very close,
as suggested in BO Cet \citep{kat21bocet}, both effects --
tidal truncation and the 3:1 resonance might work
at the same time.  If the 3:1 resonance is strong enough
and wins the competition, it may develop and eventually
cause a superoutburst.
If this fails, the behavior would look like an ordinary
IW And phenomenon.  Although this is not beyond
speculation, further observations and theoretical studies
are needed keeping this possibility in mind.
 
\subsection{Suggested observations}

   This 2022 very unexpected phenomenon was not detected
in real time, but was found during an inspection of public
archives.  This was primarily due to the lack of regular
monitoring and reporting of this object to VSOLJ or VSNET
despite its brightness (14--15~mag), well withing reach
of many amateur instruments.  This probably came from
the old impression of BO Cet -- an NL-type star with little
variation and a dull target for nightly monitoring.
This star never gained popularity during the history
of VSNET: only sporadic observations were reported
by observers who are mainly interested in NL-type stars.
Please remember that the behavior of BO~Cet completely changed
after 2019 and it should deserve more regular attention
as for other dwarf novae already on visual/CCD observers' menus.
Such a change in state in a CV may be a transient one
and may not easily be reproduced,
just as we have seen in the case of BK Lyn, which had been
known as an NL star but showed a transient ER UMa-type
state \citep{kem12bklynsass,pat13bklyn,Pdot4,Pdot5}.
This ER~UMa-type state (probably had lasted for a decade)
in BK Lyn ended in 2013 and has not repeated again.

   Due to the long orbital period (compared to typical
SU UMa stars), occasional observations to search for
superhumps may not be very attracting since it requires
relatively long time.  If your telescope can point
the object quickly, several snapshots per night would be
helpful since a small number of nightly observations by
ATLAS and ASAS-SN detected superhumps, a more organized
observations should obtain a better result.  More intensive
time-resolved photometry sessions will clarify the presence
or absence of superhumps during particular outburst --
when the object is brighter than 14.0 mag, the result
would be promising.

\section*{Acknowledgements}

This work was supported by JSPS KAKENHI Grant Number 21K03616.

The author is grateful to the ATLAS and ASAS-SN teams
for making their data available to the public.

This work has made use of data from the Asteroid Terrestrial-impact
Last Alert System (ATLAS) project. The Asteroid Terrestrial-impact
Last Alert System (ATLAS) project is primarily funded to search for
near earth asteroids through NASA grants NN12AR55G, 80NSSC18K0284,
and 80NSSC18K1575; byproducts of the NEO search include images and
catalogs from the survey area. This work was partially funded by
Kepler/K2 grant J1944/80NSSC19K0112 and HST GO-15889, and STFC
grants ST/T000198/1 and ST/S006109/1. The ATLAS science products
have been made possible through the contributions of the University
of Hawaii Institute for Astronomy, the Queen's University Belfast, 
the Space Telescope Science Institute, the South African Astronomical
Observatory, and The Millennium Institute of Astrophysics (MAS), Chile.

\section*{List of objects in this paper}
\xxinput{objlist.inc}

\section*{References}

We provide two forms of the references section (for ADS
and as published) so that the references can be easily
incorporated into ADS.

\renewcommand\refname{\textbf{References (for ADS)}}

\newcommand{\noop}[1]{}\newcommand{\hyphalt}{-}

\xxinput{bocet2022aph.bbl}

\renewcommand\refname{\textbf{References (as published)}}

\xxinput{bocet2022.bbl.vsolj}


\begin{thebibliography}{}

\bibitem[{Bruch}(2017)]{bru17CVphot1}
  {Bruch}, A.\ 2017, New\ Astron., 52, 112

\bibitem[{Fernie}(1989)]{fer89error}
  {Fernie}, J.~D.\ 1989, PASP, 101, 225 (https://doi.org/10.1086/132426)

\bibitem[{Hameury} and {Lasota}(2014)]{ham14zcam}
  {Hameury}, J.-M., \& {Lasota}, J.-P.\ 2014, A\&A, 569, A48 (arXiv:1407.3156)

\bibitem[{Hirose} and {Osaki}(1990)]{hir90SHexcess}
  {Hirose}, M., \& {Osaki}, Y.\ 1990, PASJ, 42, 135

\bibitem[{Kato}(2019)]{kat19iwandtype}
  {Kato}, T.\ 2019, PASJ, 71, 20 (arXiv:1811.05038)

\bibitem[{Kato} et~al.(2013)]{Pdot4}
  {Kato}, T., {et~al.}\ 2013, PASJ, 65, 23 (arXiv:1210.0678)

\bibitem[{Kato} et~al.(2014)]{Pdot5}
  {Kato}, T., {et~al.}\ 2014, PASJ, 66, 30 (arXiv:1310.7069)

\bibitem[{Kato} et~al.(2010)]{Pdot2}
  {Kato}, T., {et~al.}\ 2010, PASJ, 62, 1525 (arXiv:1009.5444)

\bibitem[{Kato} et~al.(2021)]{kat21bocet}
  {Kato}, T., {et~al.}\ 2021, PASJ, 73, 1280 (arXiv:2106.15028)

\bibitem[{Kato} et~al.(2004)]{VSNET}
  {Kato}, T., {Uemura}, M., {Ishioka}, R., {Nogami}, D., {Kunjaya}, C., {Baba},
  H., \& {Yamaoka}, H.\ 2004, PASJ, 56, S1 (arXiv:astro-ph/0310209)

\bibitem[{Kemp} et~al.(2012)]{kem12bklynsass}
  {Kemp}, J., {et~al.}\ 2012, in Proc. 31st Annu. Conf., Symp. on Telescope
  Science, ed. B.~D. {Warner}, \& {et~al.}  (Rancho Cucamonga: Society for
  Astronomical Sciences),  p.~7

\bibitem[{Kimura} et~al.(2020a)]{kim20kic9406652}
  {Kimura}, M., {Osaki}, Y., \& {Kato}, T.\ 2020a, PASJ, 72, 94
  (arXiv:2008.11328)

\bibitem[{Kimura} et~al.(2020b)]{kim20iwandmodel}
  {Kimura}, M., {Osaki}, Y., {Kato}, T., \& {Mineshige}, S.\ 2020b, PASJ, 72,
  22 (arXiv:1912.07217)

\bibitem[{Kochanek} et~al.(2017)]{koc17ASASSNLC}
  {Kochanek}, C.~S., {et~al.}\ 2017, PASP, 129, 104502 (arXiv:1706.07060)

\bibitem[{Lubow}(1991)]{lub91SHa}
  {Lubow}, S.~H.\ 1991, ApJ, 381, 259 (https://doi.org/10.1086/170647)

\bibitem[{Murray} et~al.(2000)]{mur00SHintermediateq}
  {Murray}, J., {Warner}, B., \& {Wickramasinghe}, D.\ 2000, New\ Astron.\
  Rev., 44, 51

\bibitem[{Osaki}(1989)]{osa89suuma}
  {Osaki}, Y.\ 1989, PASJ, 41, 1005

\bibitem[{Osaki}(1996)]{osa96review}
  {Osaki}, Y.\ 1996, PASP, 108, 39 (https://doi.org/10.1086/133689)

\bibitem[{Patterson} et~al.(2013)]{pat13bklyn}
  {Patterson}, J., {et~al.}\ 2013, MNRAS, 434, 1902 (arXiv:1212.5836)

\bibitem[{Rodr{\'\i}guez-Gil} et~al.(2007)]{rod07newswsex}
  {Rodr{\'\i}guez-Gil}, P., {Schmidtobreick}, L., \& {G{\"a}nsicke}, B.~T.\
  2007, MNRAS, 374, 1359 (arXiv:astro-ph/0611829)

\bibitem[{Shappee} et~al.(2014)]{ASASSN}
  {Shappee}, B.~J., {et~al.}\ 2014, ApJ, 788, 48 (arXiv:1310.2241)

\bibitem[{Shingles} et~al.(2021)]{shi21ALTASforced}
  {Shingles}, L., {et~al.}\ 2021, Transient Name Server AstroNote, 7, 1

\bibitem[{Stellingwerf}(1978)]{PDM}
  {Stellingwerf}, R.~F.\ 1978, ApJ, 224, 953 (https://doi.org/10.1086/156444)

\bibitem[{Tonry} et~al.(2018)]{ATLAS}
  {Tonry}, J.~L., {et~al.}\ 2018, PASP, 130, 064505 (arXiv:1802.00879)

\bibitem[{Warner}(1995)]{war95book}
  {Warner}, B.\ 1995, Cataclysmic Variable Stars
 (Cambridge: Cambridge University Press)

\bibitem[{Whitehurst}(1988)]{whi88tidal}
  {Whitehurst}, R.\ 1988, MNRAS, 232, 35
  (https://doi.org/10.1093/mnras/232.1.35)

\end{thebibliography}
\end{document}